# AirHopper: Bridging the Air-Gap between Isolated Networks and Mobile Phones using Radio Frequencies

Mordechai Guri[1], Gabi Kedma[2], Assaf Kachlon[3], Yuval Elovici[4]*

Department of Information Systems Engineering, Ben-Gurion University
*Telekom Innovation Laboratories at Ben-Gurion University

## Abstract

Information is the most critical asset of modern organizations, and accordingly coveted by adversaries. When highly sensitive data is involved, an organization may resort to air-gap isolation, in which there is no networking connection between the inner network and the external world. While infiltrating an air-gapped network has been proven feasible in recent years (e.g., Stuxnet), data exfiltration from an air-gapped network is still considered to be one of the most challenging phases of an advanced cyber-attack.

In this paper we present "AirHopper", a bifurcated malware that bridges the air-gap between an isolated network and nearby infected mobile phones using FM signals. While it is known that software can intentionally create radio emissions from a video display unit, this is the first time that mobile phones are considered in an attack model as the intended receivers of maliciously crafted radio signals. We examine the attack model and its limitations, and discuss implementation considerations such as stealth and modulation methods. Finally, we evaluate AirHopper and demonstrate how textual and binary data can be exfiltrated from physically isolated computer to mobile phones at a distance of 1-7 meters, with effective bandwidth of 13-60 Bps (Bytes per second).

[1] gurim@post.bgu.ac.il
[2] gabik@post.bgu.ac.il
[3] assafka@post.bgu.ac.il
[4] elovici@bgu.ac.il

# 1. Introduction

Modern organizations rely on their computer networks to deliver all kinds of information. Consequently, such networks are a lucrative target for malicious adversaries. Defending a network is quite a complicated task, involving host-level protection, network-level protection, secured gateways and routers, and so on. However, as long as the local area network has a connection with the outer world (e.g., the Internet), irrespective of the level of protection, an innovative and persistent attacker will eventually find a way to breach the network, to eavesdrop, and to transmit sensitive data outward.

Accordingly, when a network is used to deliver highly sensitive data, or to connect computers that store or process such data, the so called 'air gap' approach is often applied. In an air-gapped network, there is no physical networking connection (wired or wireless) with the external world, either directly or indirectly through another, less secure network. Nevertheless, employees may carry their mobile phones around the organization's workplace, possibly within the vicinity of a classified computer which is part of an air-gapped network. This may occur despite reasonable security procedures, and does not require malicious intent on the employee's part.

This scenario becomes relevant and increasingly common since a growing portion of modern mobile phones are being marketed with FM radio receivers [1] [2] [3]. With appropriate software, compatible radio signals can be produced by a compromised computer, utilizing the electromagnetic radiation associated with the video display adapter. This combination, of a transmitter with a widely used mobile receiver, creates a potential covert channel that is not being monitored by ordinary security instrumentation.

In this paper we show how mobile phones with FM receivers can be used to collect emanating radio signals, modulated with sensitive information, from a compromised computer. Although the attack model is somewhat complicated, it is not beyond the capabilities of a skilled persistent attacker. In reward for his or her effort, the attacker may collect valuable information from air-gapped networks.

## 1.1  FM Receiver in Mobile Phones

FM radio receivers in mobile phones, first introduced around 1998, have grown in popularity with the emergence of modern smartphones. Overriding early hesitations, most major carriers and manufacturers are integrating FM receivers in their products. Microsoft has endorsed FM, announcing support in an update to Windows Phone 8 [4]. Of the nearly 700 million phones sold by Nokia in 2012 and 2013, the two bestselling models (Lumia 520 and Lumia 920) now support FM radio. In the Android market, over 450 models support FM radio [1] [2]. FM receivers are also integrated in many feature phones (low-end mobile phones) and media players (e.g., iPod nano).

Additionally, legislative efforts have been made in the US to enforce integration of FM receivers in mobile phones, for use in emergency situations [5] .

## 2. Attack Model

The attack model posits that hostile code can be installed and run (a) on the targeted network, consisting of one computer or more and (b) on mobile devices that may reside in the vicinity of the targeted system. With the current popularity of 'Bring Your Own Device' (BYOD), mobile phones are often carried into and out of the physical perimeter of the organization.

The attack is composed of four main steps: (1) inserting hostile code into the target system; (2) infecting mobile phones with hostile code; (3) establishing a Command and Control (C&C) channel with the mobile phone; (4) detecting emanated signals and transmitting them back to the attacker (Figure 1).

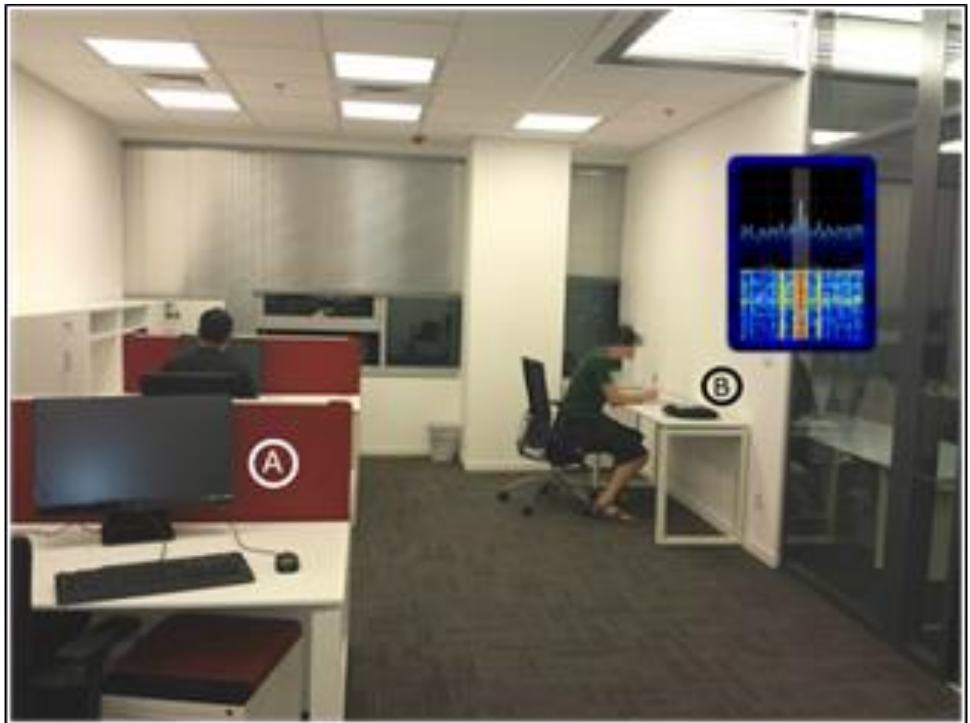

Figure 1: Demonstration of AirHopper. The targeted computer (A) emanates radio signals that are picked up by a mobile phone (B).  In this demonstration, the distance between (A) and (B) is 4 meters.

## 2.1  Infecting the Target Machine

Once the air gap barrier is breached, hostile code can infect systems within the targeted network. Such breaches are possible, as demonstrated by Stuxnet [6] [7] and Agent.btz [8]. It would be reasonable to assume that most similar cases are unpublished.  Possible attack vectors are (a) through removable media or (b) through outsourced software or hardware components. Once the hostile code runs on one node of the target network, it may propagate further and infect other nodes.

## 2.2  Infecting the Mobile Phone

Regarding mobile phones, the array of attack vectors is considerably wider than those target computers offer since these devices are connected to the outer world through a host of interfaces, such as cellular networks, Bluetooth, Wi-Fi, and physical connections to other devices. Most probably, the attacker would try to locate and infect mobile phones of those employees in the targeted organization, who are most likely to work in the vicinity of targeted computers. Similar to the early stage of an Advanced Persistent Threat (APT), this step may utilize data mining, social networks, phishing and similar social engineering methods. The mobile phone can be infected via the Internet, by visiting contaminated sites, by email messages with malicious attachments, or by installing malicious applications. Other infection routes include physical access, SMS vulnerabilities [9], and interference with the signaling interface [10].

## 2.3  Establishing the C&C Channel

After infecting the mobile phone, the attacker should establish a C&C channel. This can be done via the Internet (data or Wi-Fi), or by SMS, where messages can be sent and received without alerting the user. If suitable connectivity with the attacker is limited or discontinuous, the malicious program that runs on the mobile phone may store the accumulated information, and send it back to the attacker when the desired connection is available.

## 2.4  Radio Monitoring

The mobile phone should monitor the radio channel in order to detect emanated information. When an emanated broadcast is detected, the device decodes it. Monitoring may be: (1) continuous as long as the mobile phone is active; (2) temporal, based on the work schedule of the user; (3) spatial, based on the current location of the user, as indicated by GPS, wireless network cells, or otherwise; or (4) initiated upon receiving a remote command through the C&C channel.

## 3. Background

Generating radio signals from video card has been discussed in several works and technical papers [11] [12] [13] [14]. The main parameter required for transmission calculation is the pixel clock. The pixel clock is measured in MHz, and it represents the frequency at which the pixels are transmitted to screen such that a full frame of pixels fits within one refresh cycle. It can be calculated with a simple multiplication:

$$\left(H_{pixel} + H_{sync}\right)\left(V_{pixel} + V_{sync}\right)(Rr) = PC$$

where $H_{pixel}$ and $V_{pixel}$ are vertical and horizontal resolution (e.g., 1024 and 768) settings, $V_{sync}, H_{sync}$ are the vertical and horizontal synchronization length (in pixels) , Rr is the refresh rate (e.g., 60Hz) and PC is the pixel clock. These parameters are determined by the display standard. Generally, in both analog and digital standards, pixels timing is mainly determined by the display resolution and refresh rates.

Given the graphic card standards, it is possible to construct an image pattern that while being sent to the display will generate a carrier wave modulated with a data signal. The first option is to directly affect the voltage of the signal by changing pixel colors. Amplitude modulation (AM) is the obvious choice for data transmission of this type. By gradually changing the shade of the pixels over a fixed carrier the amplitude of the carrier signal can be changed. Previous research has demonstrated intentional generation of modulated radio signals using monitors and display adapters [15] [12] [14] . The purpose of our research is to transmit FM signals which will be received and interpreted by mobile phone FM receiver. FM signals are generated by constructing an image of alternating sequences of black and white pixels that reflect the required carrier frequency modulation. In FM modulation, white and black pixels are used to generate fixed, maximal amplitude for stronger signal generation.

## 4. Implementation

For the purpose of testing and evaluating the realistic applicability of the attack, we developed the "AirHopper" malware. In this section, we describe the implementation of the malware's two components; a transmitter implemented for the PC and a receiver implemented for the mobile phone.

Reception of the generated signals with a mobile phone's FM receiver poses new challenges and opportunities. One of the main challenges is the fact that FM-receiver chips in mobile phones are decoding the FM signals in DSP (Digital Signal Processing) chips and then generating audio signals. Consequently, the mobile phone's operating system has access only to the generated audio output (layer 2, in Figure 2) and not to carrier waves themselves. Hence, in order to send digital information to receiver, data has to be modulated using audio tones (layer 3, in Figure 2). In addition, the audio frequencies used in modulations were selected given the FM receiver characteristics and limitations, as mentioned in section 4.2.

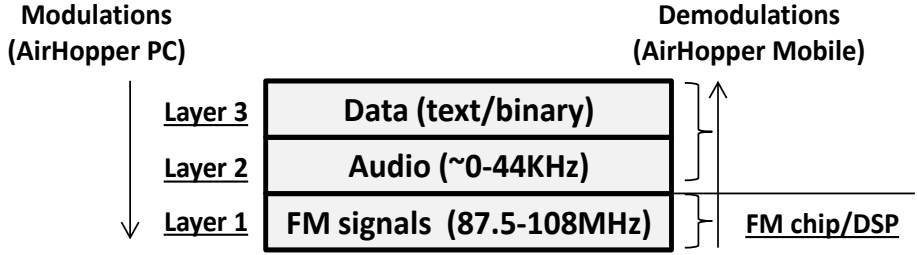

Figure 2: AirHopper modulation layers

## 4.1 FM Audio Signal Generation

In this section we describe the algorithm used by AirHopper to generate pixels pattern (image) that cause emission of specific FM audio tone. For the purpose of the audio modulation algorithm, we denote $F_c$ to be the carrier frequency which the FM receiver must be tuned to in order to receive the transmission (e.g. 90 MHz). This value is limited to be twice the pixel clock at most [15]. However, since the signal generated by the display adapter is nearly square (alternating between "black" and "white" pixels), it generates strong harmonics which may be used to ignore this limitation and transmit on a frequency higher than twice the pixel clock with some loss of signal quality [15].

$F_d$ denotes the data frequency (audio tone) to be modulated over the carrier. $P_c$ denotes the value of the pixel clock previously described and $H_p, V_p$ are the horizontal and vertical resolution parameters, plus the front and back porches which are periods of time in which no displayed pixel data is sent. This time period is used to allow for a CRT monitor time to retract its beam back to the top at the end of a frame and to the left at the end of a line. These are left over in the standard for compatibility with CRT display technology.

The modulation process outlined by the following algorithm (Algorithm 1), is loosely based on the code from Tempest for Eliza [14]. A modified variant was implemented in order to improve quality of the transmitted signal.

Intuitively, the above algorithm decides which pixel to color by emulating the carrier frequency's peaks. The resulting image is made of alternating horizontal "stripes" consisting of either a black and white pixel pattern that matches the carrier frequency or an empty black stripe. The inner loop calculates the pattern that emulates the carrier frequency $F_c$. The width of the stripes, and consequently the number of stripes is determined by the value of $F_d$ and calculated in the outer loop. Simply put, this periodic change in stripes modulates the audio tone $F_d$.

Given a pixel map, the following approximate formula can be used to determine the generated signal frequency.

$$(Y_1 - Y_0)\frac{(H_{pixel} + H_{sync})}{P_c} \approx \frac{1}{F_d}$$

$Y_0$ and $Y_1$ are the vertical pixel coordinates of two consecutive pixel rows that have the same pixel pattern. This formula uses the video card timing parameters to approximately calculate the period time of the data signal.

```
01 k ← 2 * Fd / PC , t ← 0
02 all pixels ← BLACK
03   For i ← 0 to Vp
04     IF floor(t*k)is odd
05       For j ← 0 to Hp
06         IF floor(t*k) is odd
07           pixel[j][i] ← WHITE
08         t←t+1
09     Else
10       t←t+Hp
```

Algorithm 1: Pixel map for signal tone (Fd) modulation

## 4.2 AirHopper Transmitter - Data Modulation over Audio

We used two techniques to modulate digital information over the audio signals, Audio Frequency-Shift Keying (A-FSK) and Dual-Tone Multiple-Frequency (DTMF). Our evaluation shows that A-FSK is less sensitive to interferences and can transmit to longer distances compared to DTMF. DTMF on the other hand, is more efficient in terms of bandwidth and binary data transmissions.

With both techniques, the data was modulated over audio frequencies. Our experiments show that frequencies lower than 600Hz suffered from significant interference, and extensive reception testing showed that reception distance started to diminish significantly for signal frequencies greater than 11 kHz. We limited the transmission range accordingly in both methods.

**4.2.1 A-FSK**. With this data modulation method over audio each letter or character was keyed with different audio frequency. Technically, this is done by showing an image on the entire screen area to generate a single, unique signal frequency, and keeping the image static for a short time span. The receiver is then responsible for sampling the demodulated audio signal and decoding the data. Using less than 40 distinct audio frequencies, we were able to encode simple textual data – both alphabetical and numerical. This method is very effective for transmitting short textual massages such as identifiers, key-stroking, keep-alive messages and notifications. During our extensive testing this method proved to be the most resilient to interference and had the greatest reception range and the best accurate data recovery rate. Following these positive results, we attempted modulating arbitrary binary data using 256 different audio tones, resulting in frequency spacing of 11000/256≈40Hz per character. Our tests resulted in higher error ratio and ambiguous demodulations at the decoder side. The primary reason is that the digital signal processing (DSP) component in FM receivers utilizes adaptive noise filtering, equalization and other techniques, to enhance the listening experience. These methods are altering adjacent audio frequencies, unifying them or filtering them as noise. As a result of the above, we adapted DTMF modulation to encode binary data.

**4.2.2. DTMF(16x16)**. In the DTMF schema, a combination of two tones represents a byte of data. This is similar to using two hexadecimal digits. The tones were selected from a table of 16x16 frequency pairs enabling 256 different combinations. The table's columns are frequencies in the 600Hz-5000Hz range, and the table's rows are frequencies in the 6000Hz-10400Hz range. E.g., a byte containing data equivalent to the value 134 is represented by a tone in the eighth row (⌊134/16⌋=8) and a tone in the sixth column (134 mod 16=6), played simultaneously. The transmission of two modulated audio signals is done by logically splitting the screen into two halves, transmitting a different image to each half. This resulted in some loss of range and reception quality, but the tradeoff was a much higher data transfer rate.

## 4.3  AirHopper Receiver - Data Demodulation from Audio

An essential part of the malicious code on the mobile phone is the ability to record the received FM audio and decode the data or forward the recording to the attacker. The important implementation details are described in the next sections.

**4.3.1 Android FM audio output redirection**. In order to process the FM audio, its output has to be recorded or saved. Recording output from the FM radio was undocumented in Android APIs up to API18 (Android 4.3, Jelly Bean MR2). By disassembling the MediaRecorder class file (MediaRecorder$AudioSource.smali) from the Android framework, we

found how to enable FM radio recording by using AudioRecord object with audioSource=8 and sampleRateInHz=44100 [16].

**4.3.2 Audio sampling**. Modern mobile phones support a maximum audio capture sampling rate of 44.1 KHz. The Nyquist–Shannon sampling theorem states that perfect reconstruction of a signal is possible, when the sampling frequency is greater than twice the maximum frequency of the signal being sampled. Consequently, sampling can be accomplished at 20 KHz maximum frequency. 20 KHz is the highest frequency generally audible by humans, thus making 44.1 KHz the logical choice for most audio material. Using Android's recording API, the radio signals were recorded and stored in a memory buffer, using Pulse-Code Modulation (PCM) format at 16 bit per sample.

**4.3.3 Signal processing**. Fast Fourier Transform (FFT) transforms the buffer from the time domain to the frequency domain. Since there are 1024 samples per chunk, we generate a discrete function that depicts the spectrum of the recorded signal, where the amplitude value of each index represents a frequency range of 44100/1024≈43Hz. Depending on the modulation method (A-FSK or DTMF) - the spectrum is expected to contain one or two distinct amplitudes. In DTMF modulation, there should be one frequency index with distinctively high amplitude, in each half of the spectrum range (600Hz-5000Hz and 6000Hz-10400Hz). The frequency index pair is then compared to the DTMF frequency table to recover the transmitted byte.

**4.3.4 Bypassing headphone requirement**. Some phone models require that headphones should be connected to enable the user to turn on the radio. This is necessary since the headphone cable is being used as an antenna for the FM receiver chip. Without an antenna, the reception quality of the FM receiver will be poor. While it is technically possible to receive a signal without an antenna, the headphone cable requirement ensures a good user experience. We found that this limitation is implemented by the vendor at the software level, and that it can be bypassed. We disassembled the Samsung Galaxy S3 framework file (/system/framework/framework2.odx) using baksmali [17] disassembler. In the service file (FMRadioService), the mIsHeadsetPlugged variable was initialized to 'true', and the headphone check methods ("access$302), was modified to return 'true'. This tricks the application level headphone checks to pass, regardless of the true condition.

## 4.4 Transmission Protocol

AirHopper employed two modes of transmissions; raw and structured. In raw transmissions, the data stream is taken from an array of bytes and transmitted sequentially. In case of signal loss or interruptions, the receiver is unaware of it. Streaming the raw data sequentially is suitable for textual information such as key-logging and text files or for short signaling and notifications messages. Such information is valuable even when some characters are missing. The structured protocol is appropriate for binary data transmissions when it is important to identify errors and know which part of the file or data was actually received. The transmission headers include an (1) initial synchronization pulse, (2) the sequence number of the packet, (3) the size of the packet, (4) the data itself, and (5) a checksum. AirHopper receiver can easily detect which protocol is used (raw of structured) by searching for synchronization pulse.

# 5. Hiding

In order to remain hidden while transmitting, AirHopper uses various techniques for hiding both visual aspects and transmitted signal.

## 5.1 Visual Hiding

This section explains how to eliminate the visual appearance of the transmitted data on a computer screen. Three of the suggested techniques utilize standardized means of communication with an attached monitor. HDMI, DVI and VGA monitors and VDUs implement the Display Data Channel (DDC) family of protocols.

Display Data Channel Command Interface (DDC/CI) itself is a bidirectional communication protocol between computer monitor and VDU. Monitor Control Command Set (MCCS) is a known standard, developed by Video Electronics Standards Association (VESA), that specifics the set of commands for controlling monitors over the DDC/CI communication protocol.

**5.1.1 Transmitting when monitor is off.** With DDC/CI, the transmitter can determine that a monitor is off, and only then start the transmissions. A monitor goes off as a result of power-saving settings, when the computer is idle for predefined period. As soon as the transmitter detects that the monitor is back on, it stops the transmissions. Note that since the VDU constantly generates signals, and the monitor cable is used only as an antenna, the transmission is just as effective with the monitor turned off. Technically, the monitor status is determined by checking Power Control Status (address 0xe1) of DDC/CI. The transmitter continuously samples this status, polling ioctl (I/O control) in a dedicated thread. 0 indicates that the screen is powered off.

**5.1.2 Turning the monitor off.** Using DDC/CI, the transmission code can also send a command to intentionally turn off the monitor. This method can be used in the same manner as a screen-saver; the code monitors the computer for user activity and waits for a period of idle time. In case of inactivity, it will turn off the monitor and start data transmissions until a user activity, such as a mouse movement or keyboard press, is detected. It will then stop transmitting and turn the monitor back on, in the same way a normal screen-saver or power saving feature would behave. Like the previous method, turning the monitor on/off is done by writing 1 or 0 to 0xe1, using ioctl.

**5.1.3 Switching the monitor to a different computer.** Another method for hiding the displayed image from users utilizes the popular Keyboard Video Mouse (KVM) switch. This switch is a device that allows desktop basic peripherals, keyboard mouse and monitor, to be shared between multiple computers. By clicking a button on the switch or using a keyboard key combination the user switches between desktops. Since this allows the user to comfortably use multiple desktops on one office desk, it is commonly used in organizations with air-gapped networks. Using DDC/CI, a transmitter can detect when the KVM switches between desktops and start/stop the transmissions accordingly.

**5.1.1 Using secondary display output.** Another method that we tested was to force the VDU to transmit on a specific video output even though it did not detect a connected monitor. This capability can be useful when a computer, laptop or desktop, has multiple display outputs. The transmitting code can detect which video output is currently being used and redirect the transmission to an inactive output. Since in most cases only one display cable is connected, and transmitting without a display cable as an antenna requires close proximity to the receiver, this method is effective in cases when the user places his mobile phone on the desk right next to the laptop/desktop.

Another scenario is when laptop/desktops are used in conference rooms with a secondary persistent connection to a projector. The projector output is turned off or disconnected most of the time, allowing for very long transmission periods with nothing suspicious being visible. Our testing also showed that the long, unshielded cables that are usually used to connect to projectors are also ideal for transmission quality and can be received by phones in adjacent rooms.

## 5.2 Signal Hiding

Another detection risk is accidental reception of the transmitted signal. Since the system that modulates transmitted data uses audible tone frequencies, tuning a regular FM receiver to the carrier frequency used for transmission, results in a noticeable, changing tone that may alert suspecting individuals. In order to reduce the risk of such an accidental detection, we extended the phone's FM reception band to 76 MHz – 108 MHz and the transmission carrier frequency to 80MHz.

FM transmissions between 76 MHz to 87.5 MHz are only used in very few countries, such as Japan. Most FM receivers available elsewhere in the world do not allow tuning or seeking in this range. We found, however, that although the nonstandard range is actually supported by FM chips in mobile phones, the upper software layers disable it. By transmitting in the extended range, accidental detection can be avoided in most countries. On the receiving phone, we enabled reception of the extended frequency range by modifying the Samsung framework utilizing smali [17] disassembler. Technically, the framework file is located in /system/framework. We changed the class FMRadioService in the package com.android.server by setting the value of BAND_76000_108000_kHz to 1. This modification at the OS level code is unnoticeable by the phone's user. Using a nonstandard band also helps to avoid collisions with local radio station broadcasts, and allows a wider range of available transmission frequencies. Another modification that deters accidental detection involved changing the phone's frequency seeking interval from the default 200 KHz to 50 KHz. This allows the transmitter and receiver code to fine-tune to carrier frequencies that many FM receivers will not allow for an ordinary user, hence avoiding accidental detection.

## 6. Evaluation

In this section we evaluate the feasibility and the effectiveness of the proposed data exfiltration method. The main efficiency measures include the emitted signal strength, the effective broadcasting distance, the transmission speed, the transmission quality, the battery consumption, and the scanning time. The factors include the cable type, the presence of receiver antenna, the modulation method (A-FSK/DTMF), the carrier frequency, the signal delay, the transmitted data size, and the range of transmitted frequencies. Further explanation of the various efficacy measures and factors, along with explanation of their relevance, is provided below, following a short description of the experimentation setup.

### 6.1 Setup

The monitor used was a Samsung SyncMaster SA450. The operating system was Ubuntu 13.04 (raring) 64bit, kernel 3.8.0-29-generic.

The mobile phone was a Samsung GT-I9300 Galaxy S3. The Android version was Samsung stock 4.1.2. The baseband version was I9300XXEMC2. Kernel version was 3.0.31-1042335. The radio chip was Silicon Labs Si4705 FM Radio Receiver ICs. We used the standard 1.22 meters (4 ft.) stereo headphones that came with the mobile phone. Note that in this mobile phone model, the headphones are used as the FM receiver antenna.

Four types of cables were used to connect the computer with the display unit: (I) standard VGA cable, 1.52 meters (5 ft.), 30V, 15-pin male to male;

(II) standard VGA cable, 0.91 meters (3 ft.), HD15, male to female, usually used as projector extension cable; (III) HDMI cable, 0.91 meters (3 ft.), 30V. (IV) DVI cable, 1.82 meters (6 ft.), 30V. Below we refer to these cables as standard VGA, extension VGA, HDMI and DVI, respectively. **Table 1** provides relevant classifications of the cable types (i.e. Shielded/Unshielded, Analog/Digital).

Table 1: Cable Classification

| Cable | Shielding | Signal |
|-------|-----------|--------|
| std VGA | Shielded | Analog |
| ext VGA | Unshielded | Analog |
| HDMI | Shielded | Digital |
| DVI | Shielded | Digital |

## 6.2 Signal Strength

The strength of the received signal is expected to drop as the distance from the transmitter is increased. We measured the Received Signal Strength Indication (RSSI) (see Figure 3), as well as the dBm (decibel-milliwatts, see Figure 4), across varying distances. RSSI is an indicator of signal quality usually exposed by a chip's API. The RSSI scale is arbitrarily selected by the manufacturer (Broadcom in the case of the tested mobile phone) and is usually a logarithmic scale related to the peak voltage of the received signal. dBm is also a logarithmic scale based on the power of the received signal in reference to one milliwatt (power is affected by both voltage and current).

The RSSI measurement was performed by calling the getCurrentRSSI function of IFMPlayer, an internal Samsung service used by the FM radio player. dBm values were measured with an external spectrum analyzer.

The RSSI measurements appear to be correlated with the effective distance, where a measured value of 10 on the RSSI scale roughly indicates the limit of the effective distance on each of the tested cables.

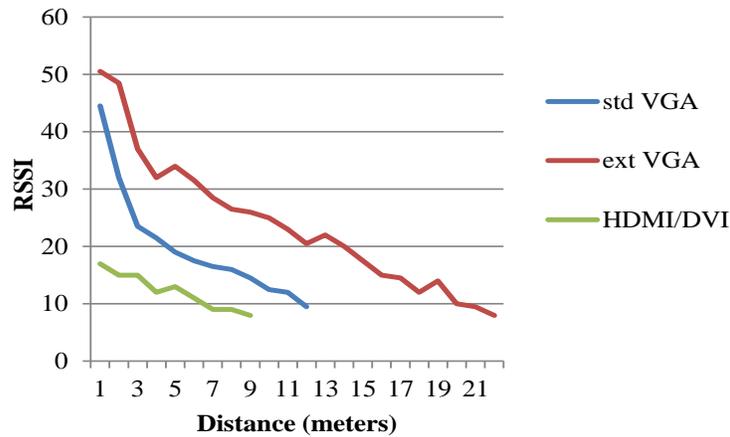

Figure 3 : Signal strength at varying distances (RSSI), using A-FSK.
(Measured by mobile-phone FM receiver)

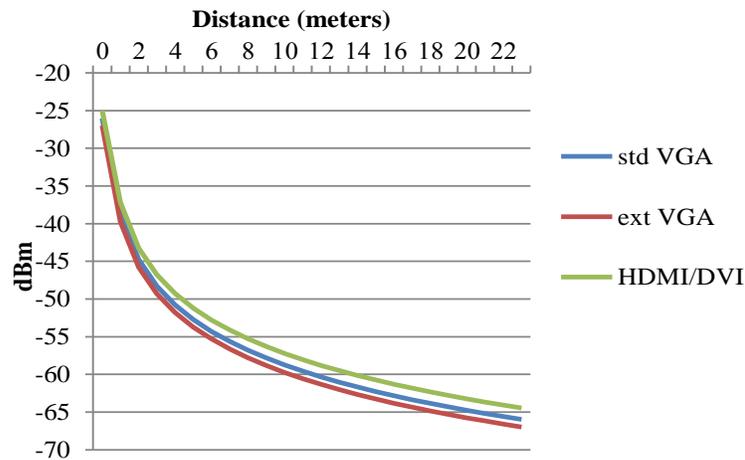

Figure 4 : Signal strength at varying distances (dBm), using A-FSK.
(Measured by an external spectrum analyzer)

## 6.3 Effective Distance

The effective distance is the maximal distance between the transmitter and the receiver, at which the transmission quality is still acceptable. As can be seen in Figure 5, the effective distance when using the receiver antenna is significantly larger with unshielded cable (extended VGA), compared to the shielded cables (HDMI and standard VGA). To summarize, with both cable types, the effective distance when the receiver antenna is present is in the order of 8-20 meters, which can be considered as useful for our purposes.

Some new models of mobile phones are equipped with built-in FM antenna, which voids the need for headphones.

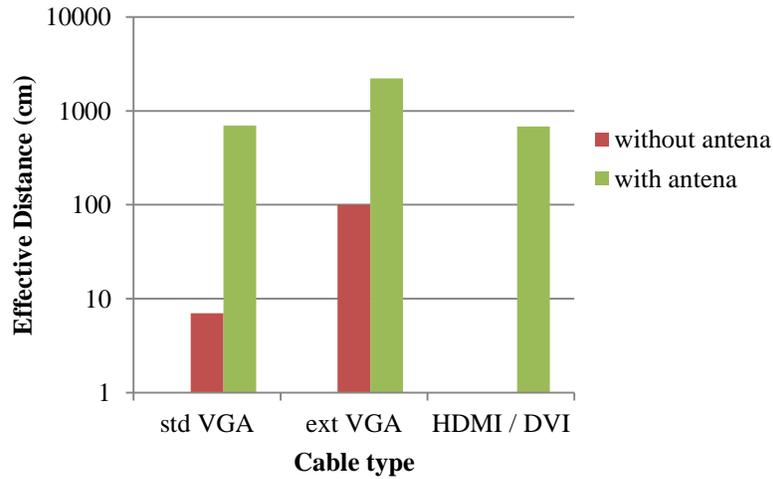

Figure 5: Signal strength at varying distances, RSSI, using A-FSK

## 6.4 Data Modulation Method

The signaling method, i.e. A-FSK or DTMF also affects the effective distance. As can be seen in Figure 6 A-FSK yields a slightly larger effective distance.

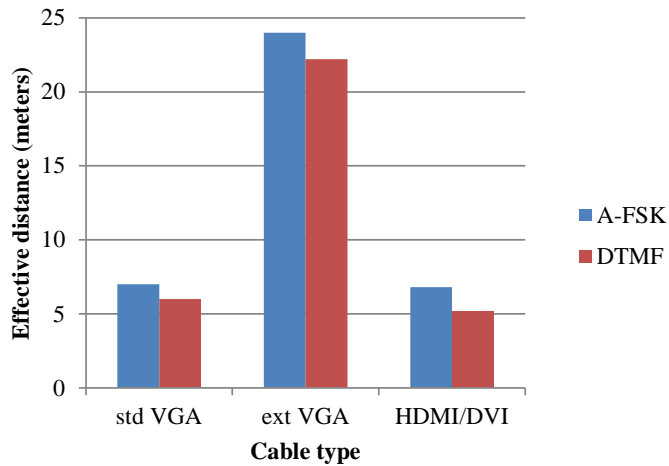

Figure 6: Effective distance with varying cable types, with receiver antenna, using A-FSK and DTMF

## 6.5 Transmission Quality

The transmission quality (also referred to as the 'success rate') is defined as the rate of correctly received bytes per originally transmitted bytes. The rate was measured against varying signal delays, i.e. the time interval (in milliseconds) assigned for transmitting one byte of data.

As can be seen in Figure 7, the correlation between the transmission quality and the signal delay is logarithmic, asymptotically approaching 100%. Beyond a signal delay of 70ms, the additional transmission quality seems negligible.

Settling on 70ms seems reasonable, since a higher delay would result in slower transmission which does not yield significant improvement in the transmission quality.

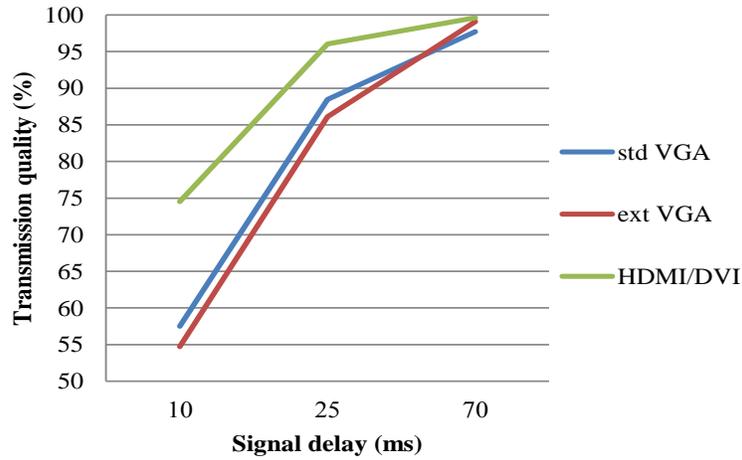

Figure 7: Transmission quality at varying signal delays, using DTMF.

As shown in **Table 2**, the transmission quality (or success rate) was measured for each cable type, over varying distances. The signal delay was 70 ms, which is the optimal value. The measurements are shown for 0.3 m, for 6.2 m, and for the maximal distance at which the transmission quality was acceptable (or the effective distance). As can be seen, the maximal distance has different values for various cable types.

The values in **Table 2** reflect the behavior of the transmission quality, which retains an almost fixed high value over an increasing distance, until it hits the maximal distance (which varies among the cable types).

Table 2: Transmission quality over varying distance
using DTMF

| Cable type | Transmission quality (%) | | | Max distance (meters) |
|---|---|---|---|---|
| | *At 0.3 m* | *At 6.2 m* | *At max* | |
| Std VGA | 99.86 | 99.27 | 97.73 | 7 |
| Ext VGA | 99.88 | 99.37 | 99.07 | 22.2 |
| HDMI | 99.21 | 99.27 | 99.61 | 6.8 |
| DVI | 99.5 | 99 | 99 | 6.8 |

## 6.6 Data Size

Effective exfiltration of data places practical limits on the transmission time. The transmitting program, running on the compromised computer, should hide its activities and has limited safe time intervals where the collected data can be broadcasted. Additionally, a receiving mobile phone does not stay within the effective range of the transmitter for indefinite time.

In **Table 3** we show the transmission time of various types of files, with typical sizes, when using raw data and when using structured packets (as described in the 'Implementation' section). A 70ms delay was found to be optimal. The time required for transmitting tiny-size data (below 100 bytes) such as IP & MAC, or password, is below ten seconds, which seems acceptable. Small-size data (below 10 KB) such as system-information or one-day keylogging would require several minutes, which also seems acceptable. However, documents in the order of 0.5 MB would take several hours to transmit and this would seem to be unacceptable. Note that the toll inflicted by using structured packets compared to raw data seems to be tolerable, and justifies using this method when accurate reception is required.

Table 3: Transmission time with varying data size (70ms delay)

| File (data size) | Raw data | Structured packets |
|---|---|---|
| IP & MAC (10 bytes) | <1 sec | 1 sec |
| Password file (100 bytes) | 8 sec | 10 sec |
| 1 hour keylogging (1 KB) | 77 sec | 106 sec |
| System information (2 KB) | 153 sec | 211 sec |
| 1 day keylogging (8 KB) | 10.25 min | 14.1 min |
| Document (0.5MB) | 10.9 hour | 15 hour |

# 7. Countermeasures

Defensive countermeasures for general Tempest threats can be categorized into technical vs. procedural. Countermeasures of the technical kind include (a) physical insulation, (b) software-based reduction of information-bearing emission, and (c) early encryption of signals. Procedural countermeasures include official practices and standards, along with legal or organizational sanctions. The relevant American and NATO standards, concerning Tempest, were held highly classified throughout the Cold War. Some of the standards eventually leaked or were released but are severely redacted [18] while more informative documents, in particular those concerning defensive countermeasures, are still classified. The prevailing standards aim at limiting the level of detectable information-bearing RF signals over a given distance on open air, or its equivalent when insulating materials are used. In effect, certified equipment is classified by 'zones' which refer to the perimeter that needs to be controlled to prevent signal reception [19]. As a countermeasure against attacks like the one in this research the 'zones' approach should be used to define the physical areas inside the organization, in which carrying a mobile phone or other radio receivers is prohibited. Another countermeasure is ensuring that only properly shielded cables are used. Shielding affects the effective range as shown by our experiments.

## 8. Related Work

Anderson [19] provides a highly informative review of emission or emanation security (Emsec): preventing attacks which use compromising emanations (CE) consisting of either conducted or radiated electromagnetic signals. In 1985, van Eck [11] demonstrated how Tempest exploits can be produced with rather ordinary equipment and skills. Using a modified TV set, he managed to reconstruct an image from electromagnetic signals produced by Video Display Unit (VDU) at a considerable distance. During the second half of the 1990s, several researches and publications related to Tempest were released [12] [13], spurring increased public interest. This trend was amplified by the Web which has offered a glimpse into classified official standards related to Tempest [18], or provided detailed instructions and tutorials related to Tempest exploits [14]. Kuhn and Anderson [12] [13] demonstrated that many of the compromising emanations from a PC can be manipulated by suitable software, in either a defensive or offensive manner. Thiele [14] [20] offers an open source program dubbed 'Tempest for Eliza', that uses the computer monitor to send out AM radio signals. The generated music can then be heard on one's radio. Kania [15] provides a collection of programs for generating FM radio signals in software, and then transmitting them using a VGA graphics card. In our implementation, we used different technique to generate FM signals in order to improve the signals quality.

## 9. Conclusion

Exfiltration of data from air-gapped networks is not a trivial task. However, in this paper we introduce AirHopper, a bifurcated attack pattern by which this challenging task can be accomplished. The core of the method consists of two essential elements: (a) electromagnetic signals emitted from a computer's monitor cable, with data intentionally modulated on those signals, and (b) an FM receiver on a mobile phone that can collect the transmitted signals, and extract the modulated data. The chain of attack is rather complicated, but is not beyond the level of skill and effort employed in modern Advanced Persistent Threats (APT). Evaluation of the proposed method includes experimental measurements of various measures such as the effective transmitting distance, cable type, the presence of receiver antenna, etc.

AirHopper adds to an understanding of electromagnetic emission threats, coupled with APT techniques. This research area is not sufficiently covered in recent academic literature. We believe that a careful professional and academic discussion of this threat, ultimately serves the interest of the cyber-defense community.